# Impact of Extreme Electrical Fields on Charge Density Distributions in Alloys

Claudia Loyola[1], Joaquin Peralta[1], Scott R. Broderick[2], and Krishna Rajan[2]

[1] Dept. of Physical Science, Universidad Andres Bello, Santiago, Chile
[2] Dept. of Materials Design and Innovation- University at Buffalo: State University of New York, USA

Abstract

The purpose of this work is to identify the field evaporation mechanism associated with charge density distribution under extreme fields, linking atom probe tomography (APT) experiments with density functional theory (DFT) modeling. DFT is used to model a materials surface bonding, which affects the evaporation field of the surface atoms under high electric fields. We show how the evaporation field of atoms is related to the charge density by comparing the directionality and localization of the electrons with the evaporation of single ions versus dimers. This evaporation mechanism is important for the reconstruction of APT data, which is partially dependent on the input evaporation fields of the atoms. In $L1_2$-$Al_3Sc$, Al-Al surface atoms are more likely to evaporate as dimers than Al-Sc surface atoms. We find that this is due to Al-Al having a shared charge density, while Al-Sc has an increased density localized around the Sc atom. Further, the role of subsurface layers on the evaporation physics of the surface atoms as a function of charge density is considered. Beyond the practical considerations of improving reconstruction of APT data, this work provides an approach for design of surface chemistry for extreme environments.

## I. Introduction

In our prior work [1], we used atomistic modeling to discriminate the evaporation field for single ion evaporations versus dimer evaporations in APT, thereby providing a more accurate input into the data reconstruction. This paper builds on that prior work by now defining the mechanism which leads to different relative bond strengths and different evaporation pathways under extreme electric field environments. By performing DFT calculations on the surface atoms and calculating the change in charge density at the surface as a function of changing electric field, we are able to correlate the bonding properties under extreme environments with the field evaporation physics.

APT operates by utilizing the physics of field evaporation, where atoms evaporate as ions from a tip with an extreme electric field due to the sharpness of the tip [2-4]. Using a data reconstruction process with some user inputs, the 3D position and chemistry of the atoms are identified [5-9]. Among these inputs is the evaporation field of the atoms. This represents potential source of uncertainty however as the atoms do not all evaporate as single ions, but rather may evaporate as dimers or trimers, with the multi-ion evaporations having different evaporation fields than the single ion evaporations. Beyond the issue of providing accurate data reconstruction inputs, the difference in evaporation fields provides information on the relative bond strengths between atoms. Therefore, identifying the difference in evaporation field between single ions and dimers provides a description of the bond strengths under extreme field environment. Further, correlating the evaporation fields to the surface chemistry provides some guidelines for design of surface chemistry. By understanding the electronic mechanism contributing to the different evaporation fields, and therefore bonding characteristics under extreme environments, we can correlate the bonding properties with chemistry and therefore improve surface design of materials for extreme conditions.

The concept of modeling the field evaporation process has been on-going for nearly half a century [10-12]. The concept of these works was then further expanded to consider the APT process [13-21]. The unique contribution of our work is the DFT calculated critical evaporation fields for surfaces as a function of chemistry and bonding mechanism. An incorrect operating evaporation field ($F_{op}$) input into the reconstruction introduces errors in the results [22-23]. As $F_{op}$ is highly dependent on the chemistry and crystal structure, we utilize a DFT approach to determine the evaporation field as a function of chemistry and structure, and furthermore to assess the relationship of these parameters on the material bonding just prior to evaporation. Of note is that a large number of the recorded multi-hit events in atom probe are not due to dimers, but just two ions recorded at the same time. However, since we are considering only relative relationships, the differences in specific ionic measurements may be considered as representing only dimers.

We further describe the field evaporation process by modeling the charge density at the surface of the material as a function of electric field, thereby identifying the charge distribution and directionality just prior to field evaporation. Further, the electron localization function (ELF) just prior to evaporation is analyzed. The remainder of the paper is organized as follows. Section 2 describes the background of the computational details used in this work; section 3 describes the acquisition of the results of the electronic charge density; section 4 correlates the charge density with APT experiments; and the electron localization functions and implications for surface design for extreme environment applications are provided in section 5.

**II. Background**

2.1 Computational Details
The computation follows the approach utilized in our prior work [1], with the calculations performed using Quantum-ESPRESSO [24] for DFT with the generalized gradient approximation (GGA). The $Al_3Sc$ cell contained 80 atoms with (111) orientation along the z-axis. A single ion or a dimer are placed at the surface, with four different configurations modeled (Al, Sc, Al-Al, and Al-Sc at the surface). The energy cutoff used is 200 Rydberg and the Marzari-Vanderbilt scheme was employed [25]. A dipole correction was used to incorporate the electric field [26]. The values identified for evaporation field from our prior work are as follows: Al – 29 V/nm, Sc – 32 V/nm,

Al-Al – 25 V/nm, and Al-Sc – 36 V/nm. The initial surface and the evaporation process which we are modeling is shown in Figure 1, for the case of Al-Sc dimer evaporation.

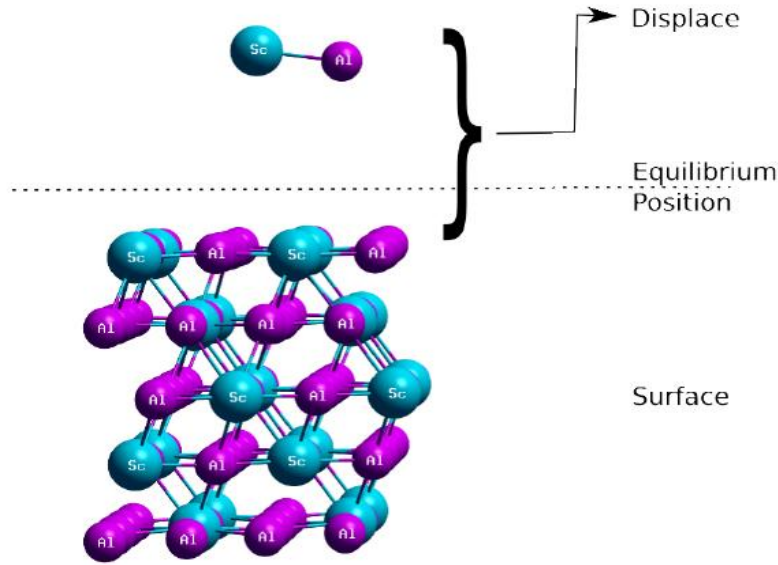

*Figure 1. Initial slab of Al$_3$Sc with A-B-C layer scheme for the surface. The evaporated ion or dimer is placed at the surface and an electric field is applied. The charge density which contributes to the evaporation of the surface ion(s), here an Al-Sc dimer, is modeled in this work. That is, correlation between surface bonding and behavior under extreme environments is assessed here.*

The electronic charge density and the Electron Localization Function (ELF) [27-28] were calculated in order to characterize the bond breaking process between the evaporating ions and the Al$_3$Sc surface when a high electric field is present. In order to isolate the role of electric field, bonding conditions with no field were also modeled. The electric field values used here correspond to just below the evaporation field (Fe) values previously calculated because the main electron densities will be present just prior to evaporation. The electric field used for Al-Al and Al-Sc ad-atoms was 22 and 33 V/nm respectively. In this procedure, the structures had been relaxed with and without an electric field. A higher grid of K-point mesh (6x6x4) in the self consistent calculation has been used to improve the electronic results.

To determine the charge density of the bond between the dimer and the surface, three different charge densities have been calculated. The first is an initial charge density $\rho0$ that corresponds to the charge density of the full system, the second is the charge density associated to the surface without the dimer $\rho 1$, and third is the charge density of only the dimer $\rho 2$. The final charge density of the bond was determined using using equation 1.

$$\rho bond = \rho 0 - \rho 1 - \rho 2. \qquad [1]$$

ELF is a mathematical function that is used to determine the properties of the bonding in a crystal between the different atomic species. The function values are defined between 0 and 1, with the value giving us information about the nature of the atomic bond. For example, a value of ELF between 0.3-0.6 is a metallic bond and 0.8 is a covalent bond. Therefore, by utilizing ELF in this work, we define not just charge density at the bond just prior to evaporation but also the nature of the bonding.

2.2 Experimental Details

The APT results shown in this paper are for an Al-3.65Mg-0.06Sc alloy. The APT used is a LEAP 3000X, with voltage mode used, flight length of 160 mm, and pressure of 5.6 x $10^{11}$ torr. The primary region of interest for this paper is the $L1_2$ Al-Sc precipitates. In this sample, the number of $Al^{2+}$ and $Sc^{2+}$ ions collected were nearly equivalent, as is shown in the mass-to-charge spectra for the sample (Figure 2).

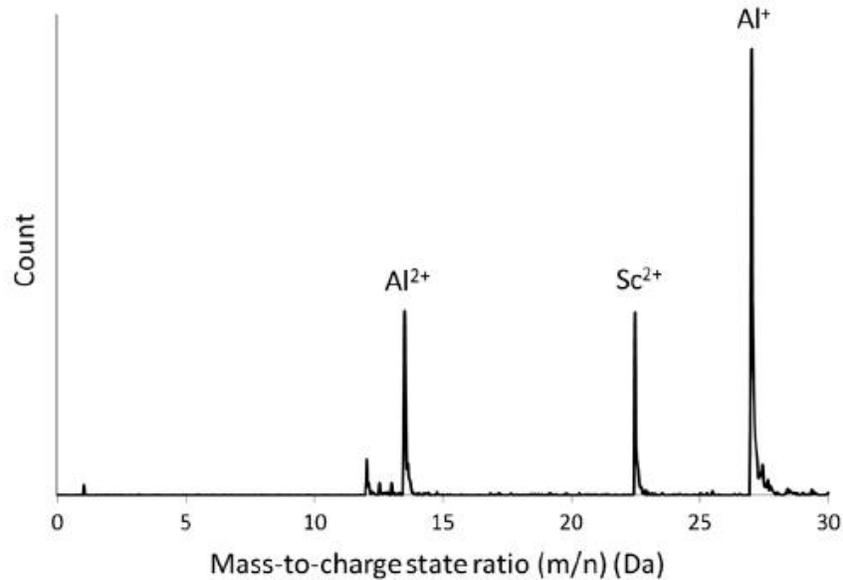

*Figure 2. Mass-to-charge spectra of the Al-Mg-Sc sample from APT. The Sc appears only in the $L1_2$-$Al_3Sc$ precipitate phase, providing a comparison with the $L1_2$-$Al_3Sc$ DFT calculations.*

## III. Charge Density Results

The effects of the electric field in the single Al and Sc atoms are presented in Figures 3 and 4, respectively. We observe a significant modification of the charge distribution that surrounds the ion when the evaporation field is incorporated. The colors for the charge densities correspond to a positive charge per volume for red and a negative charge per volume for blue. Values of +0.0015 e/bohr$^3$ and -0.0015 e/bohr$^3$ were chosen as the main representative values for observing the changes in the bonding.

The results of the charge density calculations under no electric field and with electric field just below the evaporation field are shown for Al-Al dimer and Al-Sc dimer in Figures 5 and 6, respectively. These figures present the difference of the charge density in the bonds of interest, allowing us to define the planes of interest. For the case of Al-Al dimer, we observe two principal planes of interest for the charge density. The first one P1*AlAl* corresponds to the plane that crosses the dimer atoms and one Sc atom from the surface (Sc$_s$) with which the dimer atoms are bonded. We find that they share charge density. The second plane P2*AlAl* is defined by one Al atom of the dimer and two atoms from the surface, Sc$_s$ and Al$_s$. The bond of these atoms corresponds to a significant charge density.

For the Al-Sc dimer on the surface, we can observe principal planes of interest associated to the charge density between the dimer and the surface. The first plane P1*AlSc* is defined for the dimer atoms and one aluminum atom from the surface (Al1). The second plane P2*AlSc* is defined by the Al atom of the dimer and two Al atoms from the surface, Al1 and one Al(Al2) atom that it is close to the dimer and has a shared charge density with the others atoms. The last plane P3*AlSc* corresponds to the Sc atom of the dimer and two surface atoms, Al1 and Al(Al3) atom that is close to the Sc atom of the dimer. These planes of interest are extracted from the charge density and are used to calculate the ELF.

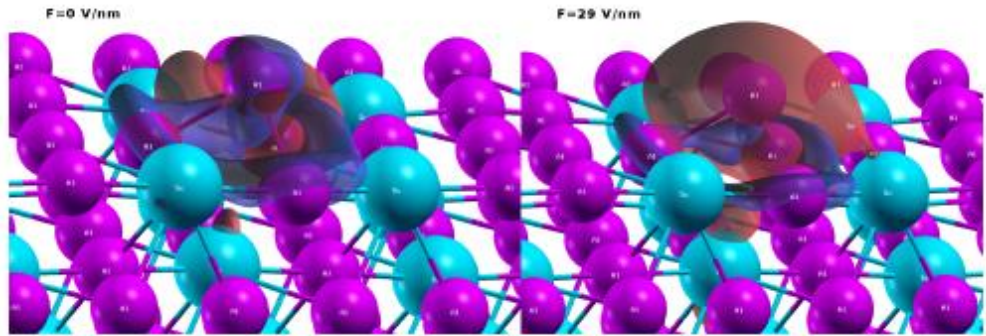

*Figure 3. Charge distribution on an Al atom on an $Al_3Sc$ without (left) and with (right) the incorporation of an electric field of 29 V/nm. The colors represent the charge densities of +0.0015 $e/bhor^3$ (red) and -0.0015 $e/bhor^3$ (blue).*

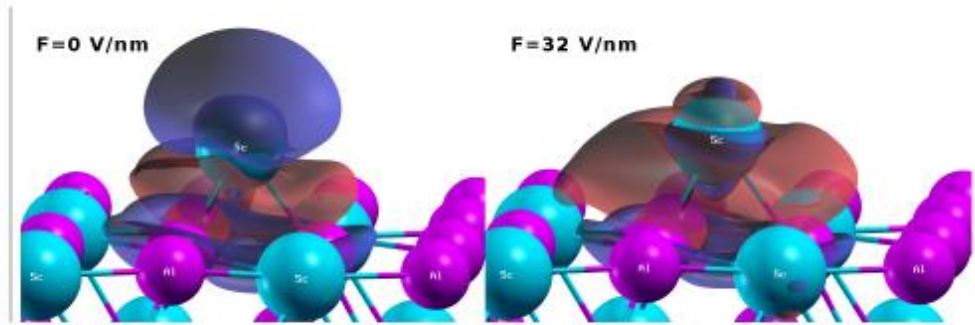

*Figure 4. Charge distribution of a Sc atom on a $Al_3Sc$ surface without (left) and with (right) the incorporation of an electric field of 32 V/nm. The differences from Figure 3 provide an insight into the different evaporation mechanisms of Al and Sc under electric field.*

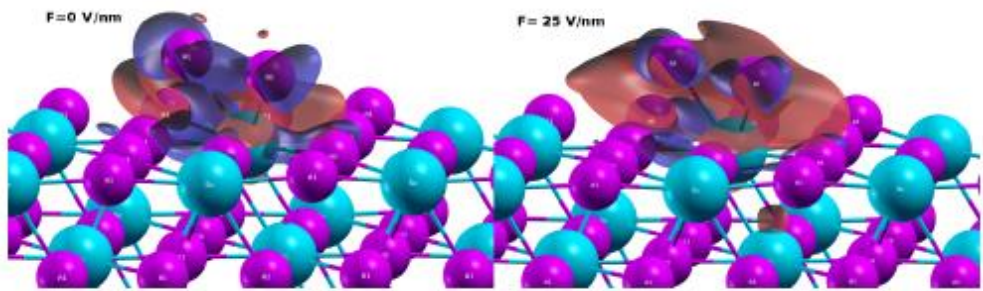

*Figure 5. Charge density for Al-Al dimer at (left) no electric field and (right) at electric field just below the evaporation field. We find that the Al dimer atoms have an even distribution of charge and that the charge is shared by the atoms. The charge is also focused between the dimer and the Sc atom ($Sc_S$). This figure defines why the evaporation field for Al-Al dimer is lower than for Al single ion, as the charge is between the dimer and the surface, with no charge in between the two atoms.*

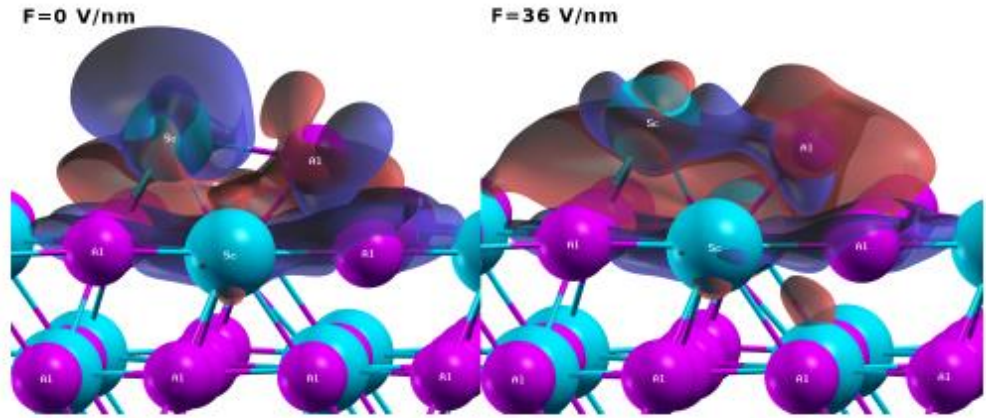

*Figure 6. Charge density for Al-Sc dimer at (left) no electric field and (right) at electric field just below the evaporation field. In this case, there is a charge localization around the Sc atom. Further, there is isolated charge between the Sc atom and the surface, and the Al atom and the surface, but these charges are not shared. This demonstrates why Sc and Al are likely to evaporate as separate ions instead of as a dimer.*

Beyond defining the critical bonding changes with increasing electric field, we also identify the difference in mechanism for evaporation. For Al-Al dimer on the surface, the primary charge is between the surface and the dimer, with the distribution shared for the dimer. This shared charge explains why Al-Al more easily evaporates in this configuration as a dimer instead of as single ions. Conversely, Al-Sc dimer on the surface has a significant charge in between the Al and Sc atoms, and also isolated charges between the atoms and the surface. This configuration of the charge density describes the mechanism for the atoms evaporating as separate ions. Therefore, by calculating the charge density, we have been able to differentiate two separate evaporation mechanisms under changing electric field.

**IV. Correlating Charge Density with APT Evaporations**

In our previous paper, we correlated the DFT calculations on field evaporation with the APT data by utilizing ion evaporation maps [1,29]. In the case of multi-hit events (that is, more than one ion detected at the same time), the ion evaporation map can be used to plot the pair-wise interactions. The axes, as shown in Figure 4, are mass-to-charge (m/n) 1 and m/n 2, where each axis represents one of the ions in a multi-hit event. The order of the ions is also inverted, so that the m/n 1 = m/n 2 line is a line of symmetry. As mentioned, many of the multi-hit events are not due to dimer evaporations, but we address this noise issue by considering only relative differences in the multi-ion events. The ion evaporation map is then correlated with relative bond strengths under extreme field, where the greater the likelihood of dimer evaporations indicating an increased bond strength. That is, it is more favorable to break all the surface bonds than to break the single bond between the dimer ions. This is clearly demonstrated in Figures 5 and 6, with the charge build-up with the surface bonds for Al-Al case, and build-up between the Al and Sc atoms in the Al-Sc case.

To correlate the DFT charge density calculations with the APT experimental data, we compare the ion evaporation map with the DFT results (Figure 7). A direct comparison between Al-Al dimers and Al-Sc dimers can be made by comparing $Al^{2+}$ (m/n = 13.5) and $Al^+$ (m/n = 27.0) with $Sc^{2+}$ (m/n = 22.5) and $Al^+$. The reason is that the number of $Al^{2+}$ hits and $Sc^{2+}$ hits in the experiment were nearly equivalent (see Figure 2). Therefore, any change in dimer concentration for these two points is due to increased number of dimers, and not a result of increased atomic concentration in the material. The obvious differences in charge density associated with Al-Al and Al-Sc dimers can be related to the evaporation fields, where surface Al-Al atoms are more likely to evaporate as a dimer than Al-Sc surface atoms, which are more likely to evaporate as two separate ions. The proposed cause of the decreased likelihood of dimer evaporations for Al-Sc is due to the charge localization around the Sc atom (charge shown as dark blue). This figure therefore correlates the evaporation mechanism with the experimental data, providing a level of physics not provided by the experimental data alone.

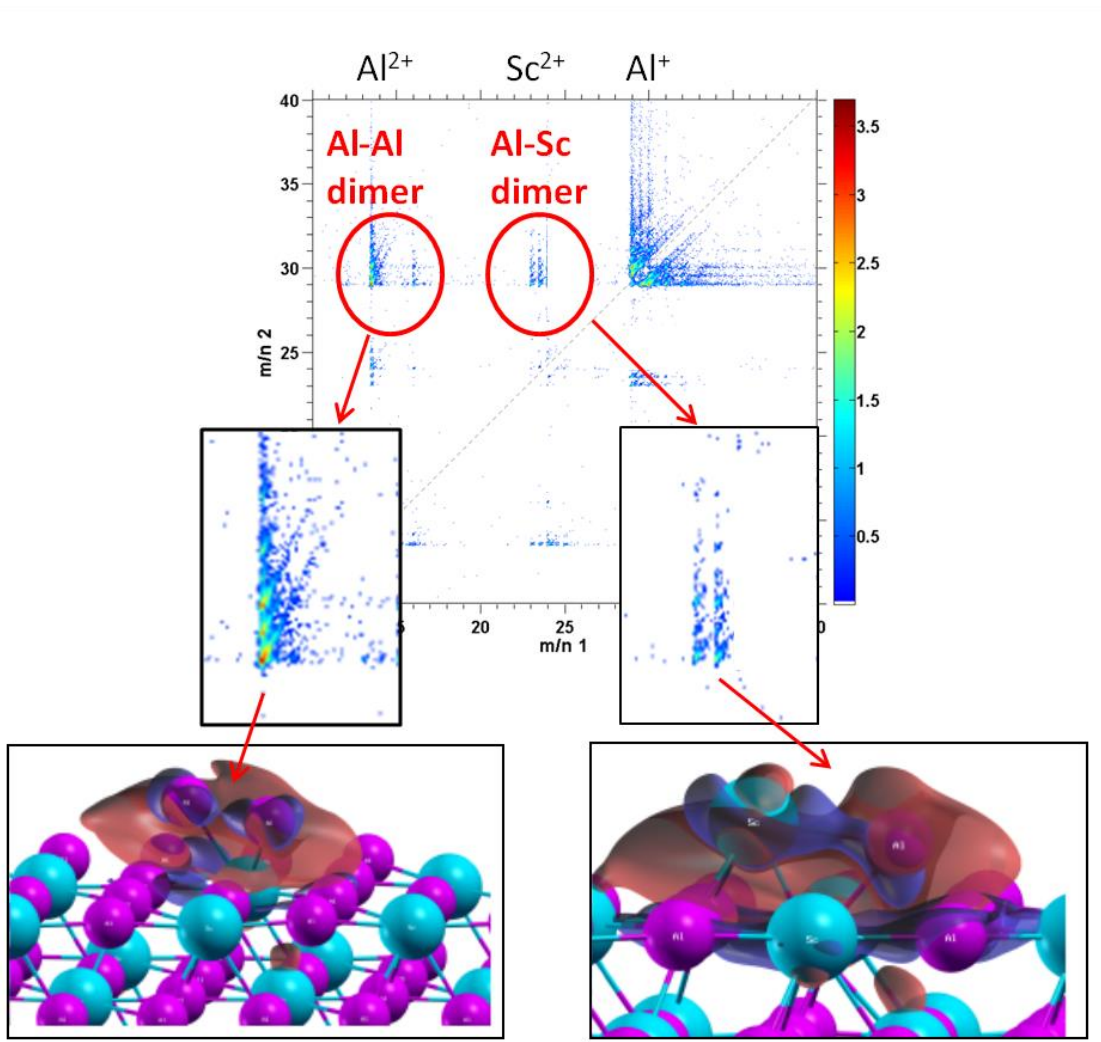

*Figure 7. Integration of APT experimental measurements with DFT data. By integrating these data we are able to include evaporation mechanisms to the data. The inset regions focus on the Al-Al dimer evaporations compared to Al-Sc dimers, with the overall chemistry of the material for these two regions being nearly equivalent. The number of Al-Al dimers is seen to be significantly higher. The DFT results describe the reason for this being the charge localization (shown as dark blue) around the Sc atom, resulting in Sc evaporating as a single ion. This figure demonstrates how DFT provides a description of evaporation mechanisms, which can then be applied to interpretation of the experimental data.*

In the case of experimental measurements of Al-Al dimers versus Al-Sc dimers, we recognize a significantly larger number of Al-Al dimers, meaning lower evaporation field for Al-Al dimers than Al-Sc dimers, as compared to single ion evaporations. The corresponding charge densities just prior to evaporation show very different evaporation mechanisms. While the charge between the Al-Sc dimer and the surface is greater than that for Al-Al, there is also much larger charge localization around the Sc atom than is seen around any of the Al atoms. We can thus propose that greater shared charge density between the surface dimers leads to increased likelihood of dimer evaporations, while localization of charge around one of the dimer ions increases likelihood of single ion evaporations.

**V. Electronic Localized Functions**
ELF provides a description of the bonding character which is not provided in the APT measurements. From the charge densities, we identified the critical planes in terms of bonding. These planes were then used for performing the ELF calculations, as shown in Figures 8 and 9 for Al-Al dimer and Al-Sc dimer, respectively. For the case of the Al-Al dimer on the surface, two different planes for the ELF have been calculated (P1*AlAl* and P2*AlAl*). For the case of the plane *P1AlAl*, the presence of electrons between the dimer atoms is incremented between a region dominated by a value of ELF of 0.8 to a region with a ELF close to 1.0 between the dimer. Another point is the narrowing of the zone of the bond between the ad-atoms and the Sc atom in the surface (Sc$_s$) where a value of 0.4 and 0.6 in the ELF is observed in both cases. For the plane *P2AlAl*, a directionality of the charge that surrounds the Al atom on the surface (Al$_s$) and the charge surrounding the Al atom of the dimer are observed.

For the case of the Al-Sc dimer on the surface, three planes P$i$*AlSc* with $i$=1,2 and 3 were used. For the plane P1*AlSc* the results show how the charge density associated to the Al dimer atom is relocated and highly concentrated in the Sc direction. ELF values close to 0.8 are located between the Al atom of the dimer and the Al atom of the surface (Al1). However, in the presence of the electric field, this value has been reduced to between 0.4 and 0.6. For the plane P2*AlSc* we can observe small changes in the charge distribution between the Al atom of the dimer and the Al atom on the surface (Al1) generating an ELF value between 0.4 and 0.6. The last plane P3*AlSc* does not show important changes, only a more clear charge close to the Sc atoms of the dimer and a more clear distribution over the Al atoms on the surface in the direction of the Sc atom.

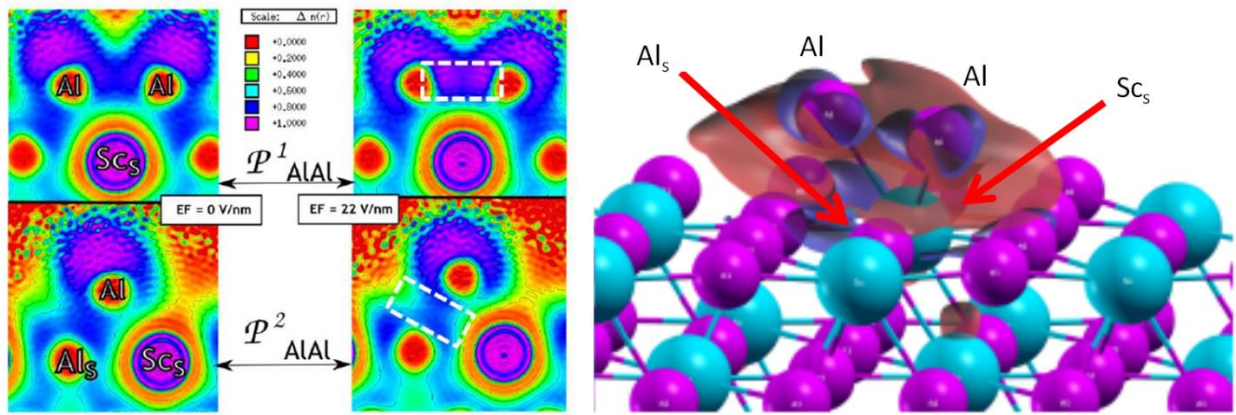

*Figure 8. The results of the ELF corresponding to the Al-Al dimer evaporation. The plane $P^1_{Al-Al}$ shows a narrowing between the Al and Sc atoms along with a high charge density between the Al atoms of the dimer. For the plane $P^2_{Al-Al}$, a similar narrowing between the Al dimer atom and the Sc atom is observed. The boxed regions show the primary regions of change with electric field, and shows that the bonding character between the dimer ions becomes more covalent with electric field.*

These results, beyond further description of the bonding, provide clear description of the change in bonding character with changing electric field. In the figures, the white boxed regions show the areas of largest change with electric field. In the case of Al-Al, we identify an increased ELF value between the Al dimer atoms with electric field, with the ELF value greater than 0.8 with electric field. This then demonstrates that the bond becomes covalent under electric field. Further, the bonding between the dimer and the surface atom decreases in ELF value to less than 0.8, so that the bonding between surface and dimer is no longer covalent under electric field conditions.

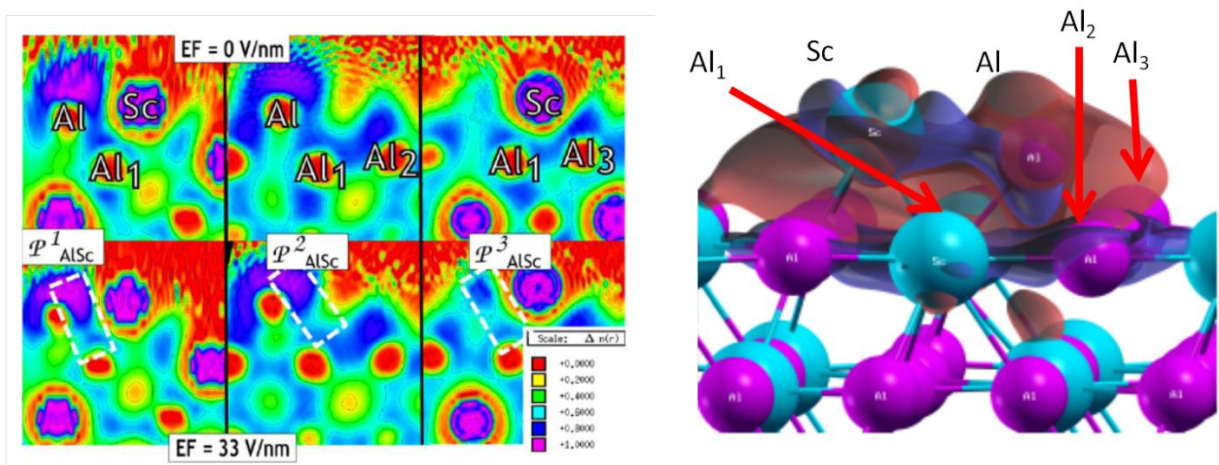

*Figure 9. The results of the ELF corresponding to the Al-Sc dimer evaporation. The plane $P^1_{Al-Sc}$ shows a charge relocation, and the plane $P^2_{Al-Sc}$ shows a distribution of the charge with increase bonding strength between the Al dimer atom and the surface. For this configuration, a clear change in the charge occurs with increasing electric field. As the Al-Sc atoms are more likely to evaporate as separate ions, this result demonstrates that increased charge activity between atoms with increasing electric field raises the barrier to dimer evaporations.*

For the Al-Sc case, we find decreased ELF values between the Al atom and the surface, with the increased ELF value isolated around the Al atom. Further, the ELF value is very low between the Sc atom and the surface. This shows that the Sc bonding is not covalent under any condition, while the Al atom and the surface have weakly covalent bonding with no electric field, and no covalent bonding character under the application of the electric field. These results for the Al-Sc case are in contrast to the Al-Al case where we identified some increase in bond strength with electric field. These calculations have therefore isolated the role of electric field on bond character in APT.

Beyond providing a mechanistic study of field evaporation, this work has applications for design of surface chemistry. That is, these calculations provide a description of bonding under electric fields, so that the design of chemistry may be considered as a function of extreme electric field conditions. The selection of surface chemistry to withstand the extreme electric fields can be selected so as to correspond with cases with increased relative bond strength and covalent character when electric field is applied. Further studies will increase the number of surface chemistry configurations to provide a larger library of possible surface chemistries which will minimize the degradation of the material under extreme field conditions. The integration of electric field with the APT experimental data contributes to bonding mechanism – electric field relationship, allowing for surface design based on the electronic reconfiguration under applied electric field.

**VI. Summary**

In this paper, DFT calculations of the charge density and bond character at a materials surface were correlated with APT experiments. The results provide the change in bonding and the mechanism for bond breaking under extreme electric field. We have also studied the charge distribution of bonds between ad-atoms and the surface to identify the evaporation pathways. Our results indicate that for $Al_3Sc$, the Al-Al dimer evaporation occurs due to a strengthening of the bond between the Al atoms in the presence of an electric field, while the Al-Sc dimer evaporation is due largely to a weakening of neighboring Al-Al bonds. However, the charge localization around the Sc atom, results in primarily single ion evaporations for the Al-Sc case. These results provide the bonding mechanism corresponding to the calculated electric fields of single ion evaporations versus dimer evaporations which improve the APT reconstruction, while also providing guidelines for surface chemistry design for materials used in extreme field applications.

**Acknowledgements**

We acknowledge support from the Air Force Office of Scientific Research (AFOSR) under grant number: FA9550-11-1-0158. SB and KR acknowledge support from AFOSR under grant number: FA9550-12-1-0456 and support from NSF under grant numbers: PHY CDI-09-41576 and ARI Program CMMI-09-389018. KR acknowledges support from the Erich Bloch Endowed Chair at the University at Buffalo-State University of New York.